# Observations of transient wave-induced mean drift profiles caused by virtual wave stresses in a two-layer system

Jan Erik H. Weber[1,†], Yiyi Whitchelo[2], Jon A. Pirolt[2], Kai H. Christensen[1,3], Jean Rabault[3] and Atle Jensen[2]

[1]Department of Geosciences, University of Oslo, PO Box 1022, Blindern, NO-0315, Oslo, Norway.

[2]Department of Mathematics, University of Oslo, PO Box 1053, Blindern, NO-0316, Oslo, Norway

[3]Norwegian Meteorological Institute, PO Box 43, Blindern, NO-0313 Oslo, Norway

An experimental study of long interfacial gravity waves was conducted in a closed wave tank containing two layers of viscous immiscible fluids. The study focuses on the development in time of the mean particle drift that occurs close to the interface where the two fluids meet. From a theoretical analysis by Weber & Christensen (*Eur. J. Mech.-B/Fluids*, vol. 77, 2019, pp. 162-170) it is predicted that the growing drift in this region is associated with the action of the virtual wave stress. This effect has not been explored experimentally before. Interfacial waves of different amplitudes and frequencies were produced by a D-shaped paddle. Particle tracking velocimetry (PTV) was used to track the time development of the Lagrangian mean drift. The finite geometry of the wave tank causes a mean return flow that is resolved by mass transport considerations. The measurements clearly demonstrate the importance of the virtual wave stress mechanism

[†]Email address for correspondence: j.e.weber@geo.uio.no

in generating wave drift currents near the interface.

## 1. Introduction

Longuet-Higgins (1953) extended the result of Stokes (1847) of mean drift in irrotational surface gravity waves (known as the Stokes drift) to fluids with viscosity. The effect of viscosity causes damping of the wave field such that some mean wave momentum is transferred to Eulerian mean currents. This transfer is achieved by the action of the virtual wave stress (VWS) at the surface (Longuet-Higgins 1969). The VWS is the manifestation of mean vorticity being generated in the oscillatory boundary layer, and it depends solely on the linear wave solutions and the rheological properties of the surface. If the surface is covered by a thin film, the VWS is greatly enhanced, and the wave-induced Eulerian mean currents become stronger than the Stokes drift in a relatively short time interval (Christensen & Weber 2005). From vorticity considerations, the increased drift in the case of an inextensible film cover was originally derived by Craik (1982).

A closed tank is typically used in a laboratory setting, with a wave maker mounted at one end, and with wave energy absorbing materials at the other end. For a one-layer system with a free surface, the VWS at any specific point will be constant in time as soon as the first wave front has traversed the full length of the tank. With a constant flux of mean momentum from the waves to the mean flow, the mean wave-induced drift currents will increase in time and space by diffusion (Longuet-Higgins 1953; Weber 2001). Eventually, a steady state will be reached when the VWS is balanced by the viscous stresses at the bottom and at the tank walls.

A similar effect occurs for interfacial gravity waves in a two-layer system; see for example Weber & Christensen (2019), hereafter referred to as WeCh. However, theoretical studies of the drift in gravity waves in a two-layer setting with different viscosities and densities are not new. In an Eulerian formulation this problem was studied by Dore (1970, 1973, 1978a, 1978b) and Wen & Liu (1995). Furthermore, Weber & Førland (1990), Piedra-Cueva (1995) and Ng (2004) applied a direct Lagrangian formulation to this problem.

Numerous experimental studies have also been conducted on drift trajectories in a two-layer system. For example, the work of Sakakiyama & Byker (1989) investigated the mass transportation in a mud layer. The work of Grue *et al.* (2000) investigated the effect of breaking and mass transport in layers of salt water with different densities, while Umeyama & Matsuki (2011) studied the trajectory of wave particles in a two-layer salt-fresh water system.

So far, the majority of the experimental studies on mass transport have been conducted with two layers of miscible fluids. As such, miscible fluids cannot maintain a sharp discontinuous density gradient at the interface (Weber *et al.* 2014), and the effect of diffusion and mixing will quickly take place in the case of large amplitude waves. This makes it difficult to measure the Lagrangian mean drift close to the interface.

For this reason, the experiments we present in this paper were performed using Isopar V produced from petroleum-based raw materials, and fresh water, constituting two immiscible fluids. The conducted experimental study aims at verifying the theoretical results by WeCh, focusing on the existence of VWS and the time development of a mean drift in interfacial waves.

There are two obvious problems that occur when we compare idealized theory with laboratory experiments. They are basically related to the finite geometry of the wave flume. First, for linear theory, the spatial damping of interfacial waves becomes larger in the tank due to the viscous effects of side-walls and rigid top and bottom planes; see e.g. the discussion by Troy & Koseff (2006). Second, when nonlinear theory is concerned, the finite geometry of the wave flume will induce a return flow opposing the Lagrangian mean drift in the wave propagation direction. We discuss this in connection with the presentation of the experimental results. The rest of the paper is organized as follows: In § 2 we recapitulate the theoretical results in WeCh, and in § 3 we describe the experimental set-up. In § 4 we present the experimental results, while § 5 compares the experimental results with theory. Furthermore, § 6 considers the effect of the VWS, while § 7 contains a discussion and some final remarks.

**2. Theory**

For details of the theory on the Lagrangian mean drift in long interfacial gravity waves, we refer to WeCh. The results in that paper are based on weak nonlinear theory for small amplitude interfacial waves in an infinitely long channel with no vertical boundaries. Upper and lower fluids are immiscible, incompressible and viscous. The waves are long compared to the depth of the two layers, and the hydrostatic assumption has been made in the vertical. Viscous effects of the upper (top) and lower (bottom) boundaries have been neglected. It is important to have these points in mind when we later compare the observed Lagrangian mean drift in a finite-length laboratory wave tank with the theory in WeCh.

In the following, we state the main results that will be useful when comparing with the laboratory measurements reported in the coming paragraphs. Analogous to the one-layer case with a free surface, the Lagrangian mean drift velocity $u_L$ in spatially damped waves is composed of a (basically irrotational) Stokes drift $u_S$, a time-independent, viscous boundary-layer solution $u_B$ (sometimes referred to as the boundary layer "steady streaming"), and an Eulerian mean current $u_E$. The last quantity develops in a diffusive manner in space and time because $u_S + u_B$ does not fulfill the boundary condition at the common interface. In this way, the initial mean wave momentum is transferred to Eulerian mean currents; see Longuet-Higgins (1953).

We consider two horizontal layers of immiscible incompressible viscous fluids. The undisturbed layer depths are $H_1$ and $H_2$ with densities $\rho_1$ and $\rho_2$ ($\rho_2 > \rho_1$), where subscripts 1 and 2 refer to the upper and lower layer, respectively. The corresponding viscosities are $\nu_1$ and $\nu_2$. We place the $x$-axis along the undisturbed interface and the $z$-axis vertically upwards. As first shown by Longuet-Higgins (1969), the condition at the boundary that drives $u_E$ can be expressed as a so-called virtual wave stress (VWS), denoted by $\tau_w$. We here concentrate on the development of the drift current near the interface. In this case, we have by definition that

$$\tau_{w1,2}/\rho_{1,2} = \nu_{1,2}\, \partial u_{E1,2}/\partial z, \quad z = 0. \tag{2.1}$$

It should be pointed out that that although the VWS is derived from quantities that are continuous at the interface (velocity and viscous stress), the VWS itself is not continuous.

The results in WeCh are derived by using Lagrangian particle-following coordinates. In the formulas for the mean flow quoted here, we can to leading order simply replace them by the independent Eulerian coordinates $(x, z)$. Thus, we have in the two layers that

$$u_{L1,2} = u_{S1,2} + u_{B1,2} + u_{E1,2}, \tag{2.2}$$

where the Stokes drift can be written

$$u_{S1} = cA^2 \exp(-2\alpha x)/(2H_1^2), \tag{2.3}$$

$$u_{S2} = cA^2 \exp(-2\alpha x)/(2H_2^2). \tag{2.4}$$

Here $c = \omega/k$ is the phase speed, where $\omega$ is the angular frequency and $k$ the wave number. Furthermore, $\alpha$ is the spatial wave damping rate, and $A$ is the interfacial wave amplitude at the horizontal position $x = 0$.

The steady streaming in the viscous boundary layers near the interface becomes

$$u_{B1} = u_{S1}Q_1[\tfrac{3}{2}Q_1 \exp(-2\gamma_1 z) - 4\exp(-\gamma_1 z)\cos(\gamma_1 z)], \tag{2.5}$$

$$u_{B2} = u_{S2}Q_2[\tfrac{3}{2}Q_2 \exp(2\gamma_2 z) - 4\exp(\gamma_2 z)\cos(\gamma_2 z)], \tag{2.6}$$

where $\gamma_{1,2} = [\omega/(2\nu_{1,2})]^{1/2}$. The order of magnitude of the viscous boundary-layer thicknesses are

$$\delta_{1,2} = 1/\gamma_{1,2}. \tag{2.7}$$

Introducing $R = 1/(1+r)$, where $r = \nu_2^{1/2}/\nu_1^{1/2}$, we can write the constants in (2.5) and (2.6) as

$$Q_1 = (H_1 + H_2)(1 - R)/H_2, \tag{2.8}$$

$$Q_2 = (H_1 + H_2)R/H_1. \tag{2.9}$$

Finally, the Eulerian mean drift currents can be written

$$u_{E1} = -u_{S1}R[rF\,\text{erfc}(X_1) - G(2\omega t)^{1/2}\text{ierfc}(X_1)], \ z \geq 0, \tag{2.10}$$

$$u_{E2} = -u_{S1}R[F\,\text{erfc}(-X_2) + G(2\omega t)^{1/2}\text{ierfc}(-X_2)], \ z \leq 0. \tag{2.11}$$

where $X_{1,2} = z/(4\nu_{1,2}t)^{1/2}$. The constants here are given by

$$F = (1+h)[(3+r)h - 1 - 3r]/(2 - 2r), \tag{2.12}$$

$$G = r(1+h)^2/(1+r), \tag{2.13}$$

where $h = H_1/H_2$.

Parameters relevant for the experiments with small amplitudes described in this paper with Isopar V ($\nu_1 = 1.44 \times 10^{-5}$ m²s⁻¹) above water ($\nu_2 = 1.12 \times 10^{-6}$ m²s⁻¹) are $H_1 = H_2 = 0.1$ m, $\omega = 2.2$ s⁻¹, $k = 7.57$ m⁻¹ and $A = 0.0102$ m. For later reference, we depict in figure 1 the transient non-dimensional Lagrangian wave drift solutions in both layers after 25, 36 and 48 s from WeCh for this set of parameters.

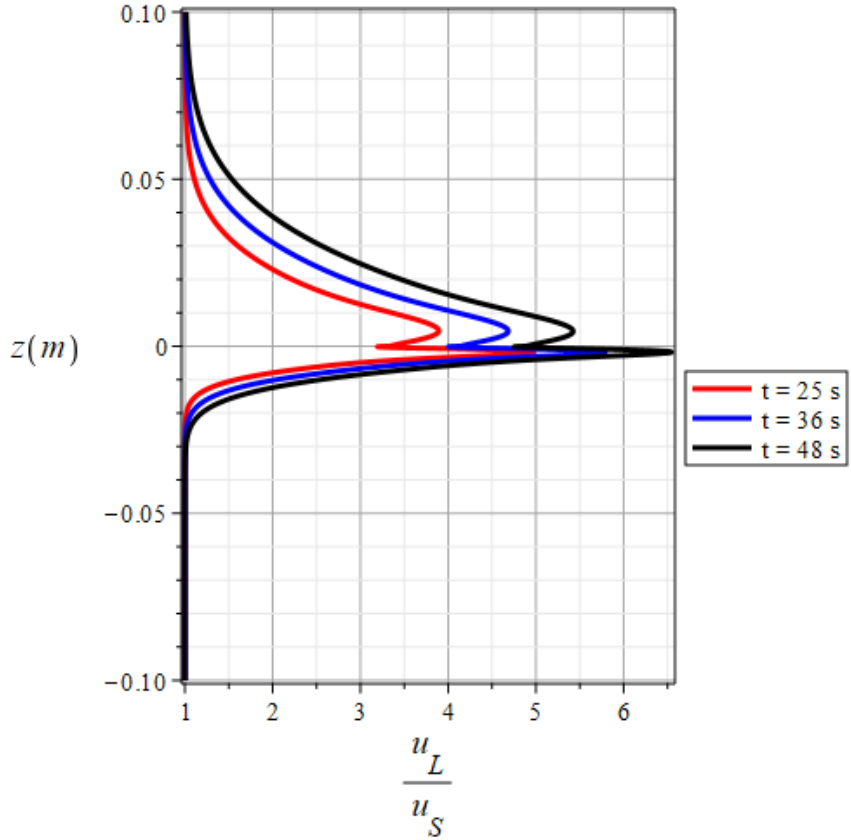

Figure 1. Non-dimensional Lagrangian wave drift solutions as function of height from WeCh at various times $t$. Here $u_{S1} = u_{S2} = u_S$.

Actually, the laboratory experiments, which we discuss later on in §4, were performed for a wider range of amplitudes and frequencies than those used to construct figure 1; see Pirolt (2021). However, for comparison with the weakly nonlinear theory in WeCh, the selected experimental amplitudes had to be as small as possible for waves in the long wave regime.

In WeCh it was demonstrated that through the action of the VWS defined by (2.1), the induced Eulerian mean currents (2.10)-(2.11) exhibited a pronounced jet-like structure near the interface, as seen in figure 1. For times larger than zero, using that $H_1 = H_2$, we find from (2.10) for the upper layer

$$u_{E1} = 2u_{S1}r\left[4(t/T)^{1/2} + r - 1\right]/(1+r)^2, \ z = 0, \qquad (2.14)$$

where $T$ is the wave period. A similar result is obtained for the lower layer. Hence, the Eulerian mean currents at the interface increases in time proportional to $(t/T)^{1/2}$.

## 3. Experimental methodology

The experiments were performed in a wave tank located in the Hydrodynamics Laboratory of the University of Oslo. The dimensions of the tank are $7.3 \times 0.25$ meters (length × width). A beach, made out of a synthetic grass material similar to what is used for sports fields, was used to dampen the waves at the end of the tank. The beach was inclined at an angle of 10 degrees and intercepts the full depth of the fluid layers. We found that the beach reflected maximum 10 percent of the incoming wave energy for the longest waves ($\lambda = 0.83$ m) and down to below 3 percent for the shorter waves. In the experiments, we have used fresh water ($\rho_2 = 998$ kg m$^{-3}$) for the bottom layer, and Isopar V ($\rho_1 = 815$ kg m$^{-3}$) for the top layer. The density of the fluids was checked using a laboratory rheometer under the environmental conditions of the experiments (temperature around 20°C). In all the following, the thickness of bottom layer was kept constant at 10 cm and the top layer at 10 cm or 20 cm. For details of the paddle design and wave generation, we refer to Pirolt (2021). Rigid lids made of Styrofoam were placed on top of the Isopar V-layer and secured in place with a steel frame clamped on the wave tank. This was done in order to suppress the generations of surface waves due to paddle motion. The model and the experimental set-up are sketched in figures 2a,b.

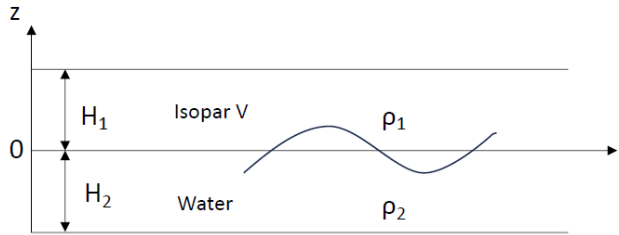

Figure 2a. Conceptual sketch of fluid model with two immiscible fluids. The lighter fluid, Isopar V, resides at the top and water at the bottom.

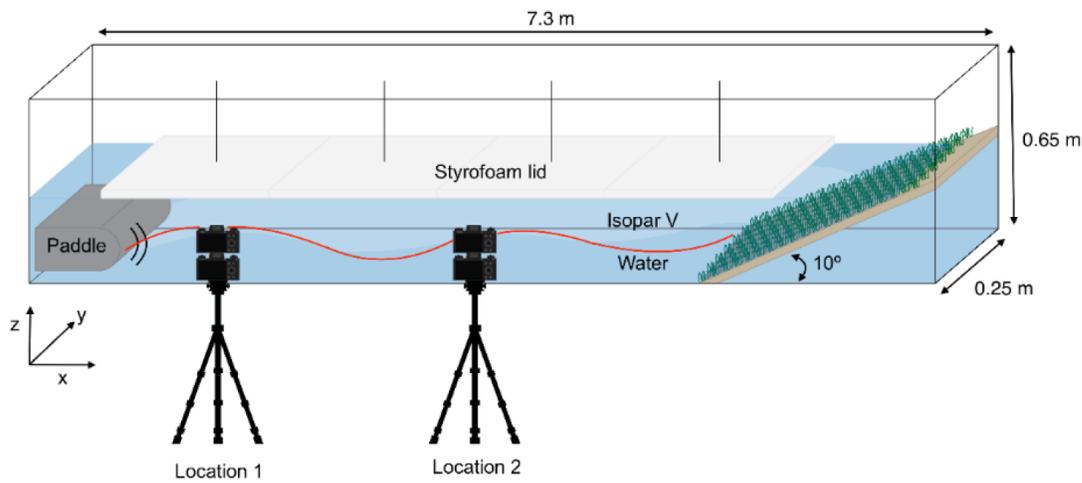

Figure 2b. Illustration of the wave tank. A half-ellipsoid wave paddle was attached at one end of the tank and a beach inclined at 10° was placed at the opposite end. Camera locations (1) and (2) were used for the PTV measurements.

Camera locations (1) and (2) were positioned at approximately 1 and 3 m distance from the wave maker. Each location was equipped with two AOS PROMON U750 MONO cameras, one upper camera for the Isopar V-layer and one lower camera for the water layer. The cameras were tilted at an angle of maximum $\pm 10°$ to avoid capturing the area where the Isopar V adhere to the glass wall of the tank (leading to blurred regions in the images) whilst also focusing on the water-Isopar V interference site. The distortion of the image captured due to this angle was considered in the "pixel to world" coordinates by using a cubic transform. Each camera has a resolution of $1280 \times 1024$ pixels, which correspond to 1 pixel = 0.01 mm, and the experimental images were captured at a frame rate of 50 fps.

The mid plane of the water column at each locations was illuminated with a LED light sheet placed below the transparent base of the tank. Each LED light sheet had a thickness of approximately 1.5 cm and it illuminated a section of the tank around 0.5 m long. The side of the tank on the opposite side of the cameras was covered with black (non-reflective) paper so the background of the lab would not interfere with the visualization of the wave motion in the tank.

For a general run, the water layer and Isopar V layers were seeded separately with polyamide particles with diameters 50 and 100 µm. The particles have a density of 1.03 g cm$^{-3}$ and were found to be almost neutrally buoyant in both fluids. The fluids were then gently stirred and sufficient time was allowed for the fluid motion to halt, securing a homogeneous particle distribution in the tank.

The paddle generated monochromatic waves that propagated down the length of the tank. The wave characteristics were measured by using image processing in MATLAB. First, we constructed a time series image from selecting the same pixel column of each image in an experimental run. The interface was identified by light reflecting from the passive particles that gathered on the surface of the water layer, forming a distinctive high luminosity line at the interface between the fluids. Then, from the time series images, a script was run to trace the middle of the interface light line and then transform the distorted instantaneous coordinates into reference world coordinates. Finally, we extracted the amplitude and frequency for each experiment.

For the PTV analysis, the images from an experimental run were first pre-processed to mask out the unused regions. Then, a PTV analysis was used to measure the instantaneous Lagrangian mean drift at the so-called end of each period, so that the interface lies at its reference height. The program uses blob tracking to identify each particle and track it for each period in DigiFlow (Dalziel 2006). The Lagrangian mean drift is then calculated by measuring the distance between the start and end of a successful trace and dividing it by the period.

The boundary-layer thickness is difficult to measure experimentally. In the experiments, $\delta_1$ and $\delta_2$ defined by (2.7) were 0.004 m and 0.001 m, respectively. One possible way is to analyze the shape of the Lagrangian mean drift from PTV tracking. However, the PTV measurements have a tendency to miss the particles very close to the interface. One reason is related to the mean $z$-position of the particle path, which is the center of the particle trajectory for a given period. Given our experimental images, the program will fail to track any particles that blurred into the interface shine by the LED light sheet. Another reason could be that the particles travel out of the field of view, leading to insufficient tracking for a whole period, which is needed for calculating the Lagrangian mean drift. Consequently, the subtle difference between the Eulerian and Lagrangian mean position of the material interface, as discussed in WeCh, could not be detected.

**4. Experimental results**

First, we give a qualitative overview of the experimental results in both layers. For this purpose, we choose data from camera location (2), with frequency $\omega = 2.2 \text{ s}^{-1}$, and amplitude $A = 1.48$ cm. Figure 3 depicts the vertical variation of the Lagrangian mean drift for successive periods in this case.

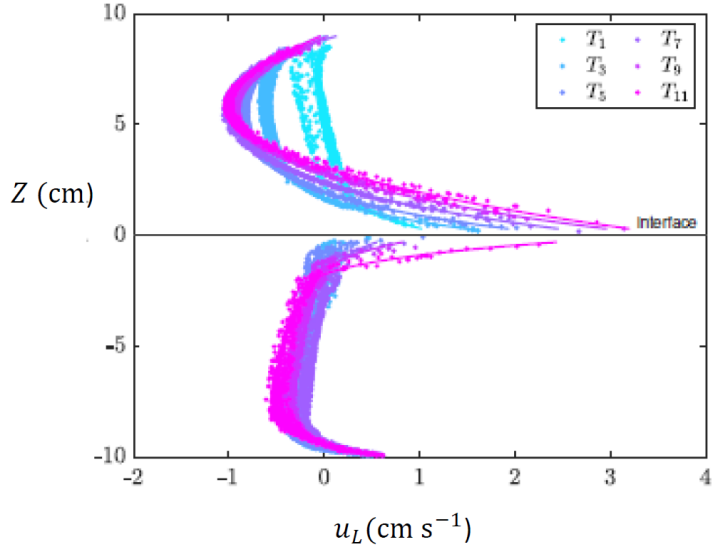

Figure 3. Lagrangian drift velocity as function of height for successive periods $T_n$. Here $H_1 = H_2 = 0.1$ m, $\omega = 2.2 \text{ s}^{-1}$, and $A = 0.0148$ m. The solid lines are the best fit to the scattered data points, yielding the Lagrangian mean flow as function of height.

We notice the qualitative resemblance between the theoretical results, as depicted in figure 1, and the experimental results in figure 3. Especially, we remark the pronounced jet-like behavior of the drift current near the interface. However, we also note the return flow in the experiments, which is not accounted for by the theory. In WeCh, it was assumed that the wave channel was infinitely long with no vertical walls that could obstruct the horizontal mean flow. In the laboratory, on the other hand, we have a finite length wave tank that is closed at both ends. Accordingly, as soon as the wave field is established throughout the length of the tank, there must be a negative mass flux balancing the positive wave-induced Lagrangian flux in each layer.

As seen in figure 3, the development of the drift current in time and space is qualitatively the same in both layers. However, it was somewhat problematic to process the data from the (lower) water layer. In many cases, we did not get close enough to the interface to observe the steady streaming, and overall, the tracking did not give a consistent, repeatable result. In conclusion, we find that the camera resolution was insufficient to produce high quality PTV tracking in the lower layer. For a quantitative comparison with the theoretical results, we therefore concentrate on the data from the (upper) oil layer, where the processing was easier.

4.1. *Upper layer measurements*

The theoretical results in WeCh are valid for small-amplitude waves, where $A/H_{1,2} \ll 1$. The experimental parameter set that comes closest to the assumptions in the theory (used to construct figure 1) is $H_1 = H_2 = 0.1$ m, $\omega = 2.2$ s$^{-1}$, $k = 7.57$ m$^{-1}$, and $A = 0.0102$ m at camera location (2). The Lagrangian measurements of the upper-layer mean drift in this case are depicted in figure 4.

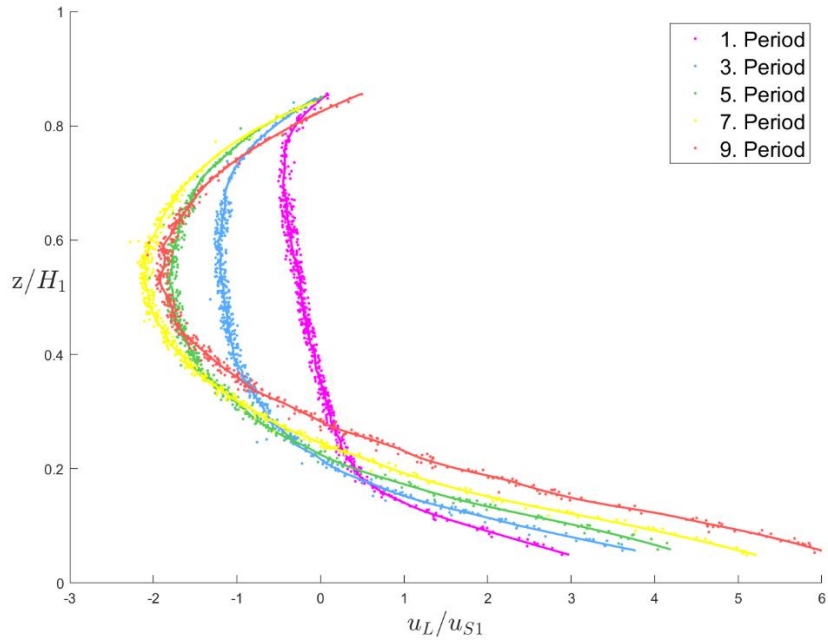

Figure 4. Experimental Lagrangian drift results for the upper layer (dotted points) for successive periods when $H_1 = H_2 = 0.1$ m, $\omega = 2.2$ s$^{-1}$, $k = 7.57$ m$^{-1}$ and $A = 0.0102$ m. The solid lines are the best fit to the scattered data points, yielding the Lagrangian mean flow as a function of height at various times.

In the regions close to the interface (inside the viscous boundary layer) and in the viscous region near the upper rigid boundary, reliable measurements could not be obtained. This explains the lack of data points in the figure.

In figure 5, we have displayed the interfacial elevation as function of time. Periods and color codes correspond to those in figure 4.

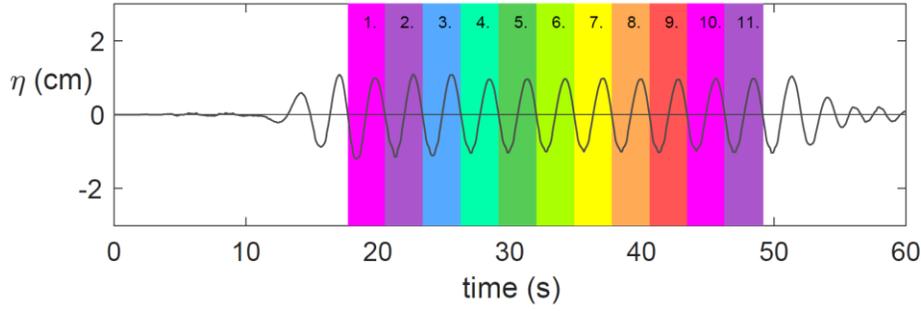

Figure 5. Time series showing chosen periods for the plotted data in figure 4. Here, $\eta$ is the interfacial elevation at camera position (2).

Again, it is particularly interesting to note the increase in time of the Lagrangian drift velocity close to the interface in figure 4. This is a strong indication that the Lagrangian drift develops in the manner predicted by the theory of WeCh; see e.g. figure 1. For a quantitative comparison, we have to adjust for the return flow in the tank, which is an obvious feature in figure 3 and 4. This will be attempted in §5.

## 5. Comparison with theory

To compare the experimental results with the theory in WeCh, we must quantify the return flow in the experiments seen in figure 3 and figure 4. As discussed in § 4, volume conservation in the upper layer requires that

$$\int_0^{H_1}(u_{L1} + u_{R1})dz = 0, \qquad (5.1)$$

where $u_R$ is the return flow. For continuity reasons, this relation must be valid at any horizontal position down the tank. The corresponding return flux becomes

$$U_{R1} \equiv \int_0^{H_1} u_{R1}\, dz = -\int_0^{H_1} u_{L1}\, dz. \qquad (5.2)$$

Obviously, this flux varies with time. As a first approximation, we take that the return flow is parabolic. This is the simplest form that yields a net transport while satisfying the top and bottom no-slip boundary conditions in the tank. The parabolic approximation has previously been observed to provide good agreement with wave tank experimental flow profiles (Rabault *et al*. 2016). From the observations, we note a return flow maximum near the middle of each layer. Hence, we assume

$$u_{R1} = K_1 z(z - H_1). \quad (5.3)$$

We then find for the time-dependent coefficient that

$$K_1 = -6 U_{R1}/H_1^3, \quad (5.4)$$

where $U_{R1}$ is given by (5.2). However, it should be remarked from the experimental data presented in figure 4 that for small times (1. period), the parabolic form of the return flow is not very well developed, while for longer times the profile shows a much more parabolic behavior.

In figure 6, we have depicted the theoretical result $u_L/u_{S1}$ from (2.2) by adding the model result $u_{R1}/u_{S1}$ from (5.3). In the same figure, we have replotted the experimental data for the Lagrangian drift shown in figure 4. The vertical plot range in the figure is limited to the region where we have reliable measurements, i.e. between the viscous boundary layers at the interface and at the top.

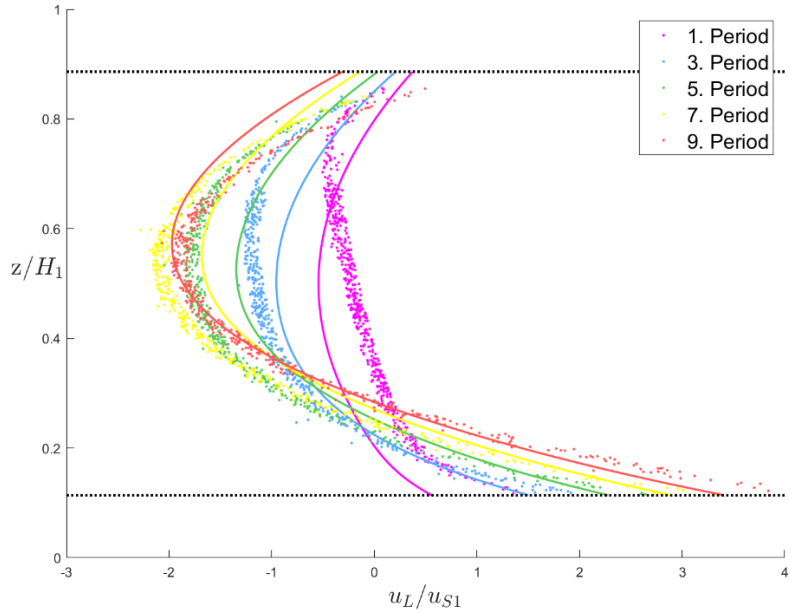

Figure 6. The experimental Lagrangian drift data from figure 4 displayed together with the theoretical result (2.2) of WeCh (solid lines), when the return flow (5.3) is added. The horizontal dotted lines mark the outer end of the viscous boundary layers (thickness $\pi\delta_1$) at the interface and at the top .

From the results presented in figure 6, we conclude that the theory in WeCh, with the added return flow, reproduces the experimental results quite well. This is particularly so for larger times when the return flow in the wave tank becomes more parabolic.

Finally, apart from the assumption of an idealized parabolic profile, it should be pointed out that the observed discrepancy between experiments and theory also could be caused by nonlinear effects not taken into account here.

## 6. Virtual wave stress

The basic intention behind the present laboratory investigation is to verify the strong transfer of mean momentum from interfacial waves to jet-like mean drift currents near the interface, as predicted by the theory in WeCh. The action of VWS at the interface is central in this theory. As explained before, we have difficulties in measuring the steady streaming $u_B$ in the viscous boundary layer, which is part of the Lagrangian velocity. But this is not crucial for the present investigation, since we are basically interested in the wave-induced Eulerian mean current, or more specifically, the spatial gradient $\partial u_E/\partial z$, which determines the VWS.

In the upper layer, we find from (2.10) that the virtual wave stress can be written

$$\tau_{w1}/\rho = \nu_1(\partial u_{E1}/\partial z)_{z=0} = u_{S1}\nu_1^{1/2}R[rF/(\pi t)^{1/2} - G\,\omega^{1/2}/2^{1/2}], \tag{6.1}$$

where the constants are given by (2.12) and (2.13). It is not possible to compare (6.1) directly with the experimental results, since we only measure the Lagrangian drift for $z \geq \delta_1$. However, outside the viscous boundary layer, we have that $\partial u_{E1}/\partial z \approx \partial u_{L1}/\partial z$. This fact can be used when we compare experimental gradients with theory.

In the measurements, the Lagrangian mean flow gradients are affected by the shear of the return flow. For a direct comparison, we must add our parabolic flow (5.3), constructed to yield zero net mass transport in the upper layer. The result is shown in figure 7, where we have compared the mean drift gradients at the edge of the viscous boundary layer.

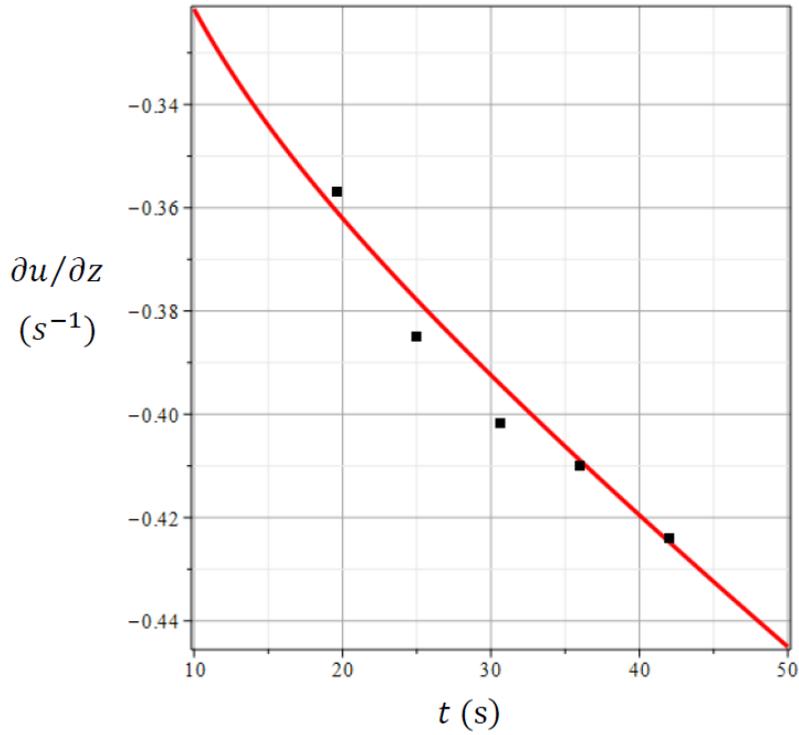

Figure 7. Theoretical Eulerian drift gradient at $z = \delta_1$ (red line) as function of time. Here $u = u_{E1} + u_{R1}$ from (2.10) and (5.3). Black squares denote value of the Lagrangian drift gradient estimated from the measurements near the interfacial boundary layer in figure 4 after $t = 19.6, \ 25, \ 30.6, \ 36,$ and $42$ seconds, corresponding to periods $1, 3, 5, 7, 9$.

We emphasize that there are uncertainties related to the determination of spatial gradients from the data, and also to the vertical position at which theory and measurements should be compared. But generally, the computed gradients are of the same order of magnitude in the theory and the measurements, and so are their variation in time. We therefore conclude that the correspondence between the theoretical and measured drift gradients seen in figure 7, demonstrates that the virtual wave stress mechanism can generate such wave drift currents as we observe near the interface in the wave tank.

## 7. Discussion and concluding remarks

We have performed laboratory experiments in a closed wave tank containing two layers of immiscible fluids. A D-shaped paddle produced interfacial waves of different frequencies and amplitudes. Four cameras, two at each measurement location in the tank, were used to capture the fluid flow motion. We used PTV measurements to determine the instantaneous Lagrangian mean drift velocity $u_L$. An increase in $u_L$ close to the interface was observed as time progressed, just as predicted by the theory of WeCh. The finite geometry of the wave tank, with impermeable walls at both ends, induces a mean horizontal return flow in both layers. By correcting for the return flow in terms of mass conservation in each layer, the theoretical and laboratory results for the Lagrangian mean flow correspond quite well. Good fit is obtained for the drift gradients outside the viscous boundary layer. This is a laboratory demonstration of the effect of the virtual wave stress in producing a jet-like particle drift near the interface in a two-layer system. To the authors' knowledge, the transient Eulerian response to the VWS has not been observed in the laboratory before.

Oil on water is commonly observed in nature, from monomolecular films to much thicker layers (Alofs & Reisbig 1972; Lange & Hühnerfuss 1978). In the latter case, the presence of oil spill is usually related to shipwrecking or accidents at offshore oil installations. Future offshore oil explorations are expected to move towards shallow Arctic regions where sea ice is present, at least for part of the year. In this case, accidental release of oil may be particularly difficult to handle in the presence of waves. Generally, ocean swell originates in the open sea and propagates towards the ice front, or the marginal ice zone; suffering increased attenuation as it moves further into the ice-covered region; see e.g. Squire & Moore (1980). As we have shown in this paper, long gravity waves on the oil/water boundary may induce a net transport oil further into the region beneath the ice. This net transport is much larger than the traditional Stokes drift. Obviously, it will complicate any oil recovery operation.

Finally, the uncertainty related to the theoretical parameterization of the return flow in a closed tank should not over-shadow the obvious need for more careful measurements near the wave generator and at the far end of the tank in order to reveal the nature of the return flow.

**Acknowledgements**

We thank Stephane Poulian, Reyna Guadalupe Ramirez de La Torre and Miroslav Kuchta for support and guidance. We also thank Olav Gundersen for assistance with the set-up of the experiments and Blandine Feneuil for information about the rheological properties of Isopar V. This work has been supported by the Research Council of Norway through Grant 280625 (Dynamics of floating ice).

**Declaration of interests**

The authors report no conflicts of interest.